# SπRIT: A time-projection chamber for symmetry-energy studies

R. Shane[1], A. McIntosh[2], T. Isobe[3], W.G. Lynch[1]*, H. Baba[3], J. Barney[1], Z. Chajecki[1], M. Chartier[4], J. Estee[1], M. Famiano[5], B. Hong[6], K. Ieki[7], G. Jhang[6], R. Lemmon[8], F. Lu[1,9], T. Murakami[10], N. Nakatsuka[10], M. Nishimura[3], R. Olsen[2], W. Powell[4], H. Sakurai[3], A. Taketani[3], S. Tangwancharoen[1], M.B. Tsang[1], T. Usukura[7], R. Wang[11], S.J. Yennello[2], J. Yurkon[1]

*corresponding author: lynch@nscl.msu.edu

[1]NSCL and Department of Physics and Astronomy, Michigan State University, East Lansing, Michigan, 48824, USA
[2]Cyclotron Institute, Texas A&M University, College Station, TX 77843, USA
[3]RIKEN Nishina Center, Hirosawa 2-1, Wako, Saitama 351 -0198, Japan
[4]Department of Physics, University of Liverpool, Liverpool, Merseyside L69 7ZE, UK
[5]Department of Physics, Western Michigan University, Kalamazoo, Michigan, 49008-5252, USA
[6]Department of Physics, Korea University, Seoul 136-703, Republic of Korea
[7]Department of Physics, Rikkyo University, Toshima -ku, Tokyo 171 -8501, Japan
[8]Nuclear Physics Group, STFC Daresbury Laboratory, Daresbury, Cheshire, WA4 4AD, UK
[9]Shanghai Institute of Applied Physics, CAS, Shanghai 201800, P. R. China
[10]Department of Physics, Kyoto University, Kita-shirakawa, Kyoto 606-8502, Japan
[11]Department of Physics, Tsinghua University, Beijing 100084, P. R. China

**Abstract**

A Time-Projection Chamber (TPC) called the SAMURAI Pion-Reconstruction and Ion-Tracker (SπRIT) has recently been constructed at Michigan State University as part of an international effort to constrain the symmetry-energy term in the nuclear Equation of State (EoS). The SπRIT TPC will be used in conjunction with the SAMURAI spectrometer at the Radioactive Isotope Beam Factory (RIBF) at RIKEN to measure yield ratios for pions and other light isospin multiplets produced in central collisions of neutron-rich heavy ions, such as $^{132}$Sn + $^{124}$Sn. The SπRIT TPC can function both as a TPC detector and as an active target. It has a vertical drift length of 50 cm, parallel to the magnetic field. Gas multiplication is achieved through the use of a multi-wire anode. Image charges are produced in the 12096 pads, and are read out with the recently developed Generic Electronics for TPCs.

## 1    Introduction

This paper describes the SAMURAI Pion-Reconstruction and Ion-Tracker (SπRIT), a Time Projection Chamber (TPC) designed for measurements of the density dependence of the nuclear symmetry energy at supra-saturation densities on the order of $\rho \approx 2\rho_0$. To achieve this goal, the TPC is designed to measure the momentum distributions of pions and isotopically resolved light particles with $Z \leq 3$ emitted in central collisions of neutron-rich nuclei, such as $^{132}$Sn + $^{112}$Sn at incident energies up to E/A=350 MeV. Such data will allow construction of experimental quantities such as the ratios of the energy spectra for pions (i.e. $\pi^-/\pi^+$) [1] and mirror nuclei such as n/p and t/$^3$He [2] that have been predicted to be sensitive to the symmetry energy at supra-saturation densities. This will provide laboratory constraints on the symmetry energy at about twice saturation density that are much needed [3]. The overall design of the SπRIT TPC is outlined in the following sections.

## 2    Detector Description

The SπRIT TPC is designed to reside over the 2 m pole face and within the gap of the Superconducting Analyzer for MUlti-particles from RAdioIsotope beams (SAMURAI) magnet at the Rare Isotope Beam Factory (RIBF) at the RIKEN Laboratory in Japan [4]. A detailed description and properties of the SAMURAI dipole can be found in Ref. [5]. To simplify the design process, some of the design parameters for the SπRIT TPC were adapted from those of the EoS TPC [6], which had similar initial magnet geometry and requirements for pion-track reconstruction. The design deviated from that of the EoS TPC to accommodate the smaller usable gap (75 cm) of the SAMURAI dipole, the dimensions of the readout electronics and to allow possible future use of the TPC as an active target.

Figure 1 shows a schematic drawing illustrating the operation of the SπRIT TPC.  The RIBF facility produces rare isotope beams, such as $^{132}$Sn, which will be directed onto a target approximately 3 mm in front of a field cage within the TPC. This beam induces a nuclear reaction in the target, producing charged particles that are largely directed into the field cage through its front window. Atoms in the P10 counter gas (90% argon, 10% methane) are ionized by these charged particles, producing secondary electrons along the ion tracks. These electrons drift in a uniform electric field produced by the field cage, pass through three wire planes and are multiplied on anode wires situated just below a segmented plane of charge-sensitive pads. The typical avalanche at the wire multiplies the number of electrons by a factor of the order of 2000. Motions of the resulting charge in the anode region produce changes in the image charges on the pads of the pad plane. These image currents are amplified and digitized by the TPC electronics, discussed in section 2.2 below. The distribution of induced charges on the pads is analyzed to determine the projection of the original charged particle's track onto the (x-z) plane of the pads. The time of arrival provides the third (y) component of the particle's trajectory. The charge per unit track length is proportional to the stopping power dE/dX along the track. This stopping power, the curvature (rigidity) and the initial direction of the track provide the particle identification and momenta of the detected particles produced in each nuclear collision. Table 1 lists some of the properties of the SπRIT TPC. Except for the drift length, constrained by the SAMURAI dipole magnet gap, they resemble those of the EoS TPC, which has a similar design [6].

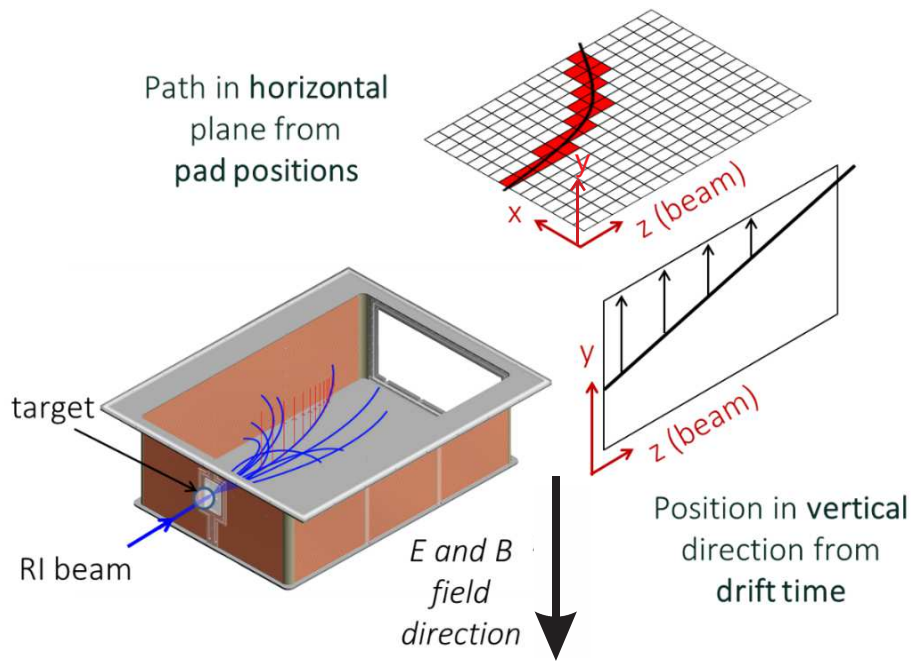

**Figure 1: Operation of TPC:** *A charged particle passing through the TPC ionizes the counter gas. The ionized electrons drift through electric field of the field cage toward a grid of charge sensitive pads, and are multiplied on anode wires near the pads. The TPC electronics amplifies the image charges induced on the pads due to the motions of charges in the avalanche region. These image charges provide a projection of the particle path on the horizontal pad plane.  The vertical component of the particles trajectory is inferred from the time of arrival.*

| SπRIT TPC Parameters | | | |
|---|---|---|---|
| **Pad Plane Area** | 1.3 m x 0.9 m | **Gas Gain** | 2000 |
| **Number of Pads** | 12096 (112x108) | **E-field** | 135 V/cm |
| **Pad size** | 12 mm x 8 mm | **Drift velocity** | 5.5 cm/µs |
| **Drift Distance** | 50 cm | **dE/dx range** | Z=1-8, π, p, d, t, He, Li-O |
| **Pressure** | 1 atmosphere | **Two-track res.** | 2.5 cm |
| **Gas composition** | 90% Ar + 10% $CH_4$ | **Multiplicity limit** | 200 |

Table 1: Relevant properties of the SπRIT TPC

The outer enclosure of the SπRIT TPC is a rectangular box, approximately 2.06m long x 1.50m wide x 0.74m high. The field cage is designed with a 50 cm drift length, parallel to the magnetic field, which is limited by the 75 cm usable gap in the dipole magnet of the SAMURAI Spectrometer magnet chamber and the space required by the read-out electronics placed on the top. The main components of the design are the top plate, pad plane, wire planes, and field cage as illustrated in the exploded view of the SπRIT TPC shown in Figure 2.

The top plate provides a rigid surface to mount the pad plane, readout electronics, and field cage. To enhance the structural rigidity of the top plate, several long "ribs" are bolted around the perimeter and across the outside surface above the pad plane. The pad plane is mounted on the inner surface, and there are 384 holes in the top plate that allow signals from the pad plane to be transmitted to the readout electronics.

The interior of the field cage is 145 cm long x 97 cm wide by 52 cm high. The side and front walls are constructed of 1.6 mm thick halogen-free G10 printed circuit boards (PCBs) with 6 mm wide copper strips and 4 mm gaps between strips corresponding to a 1 cm pitch on both the interior and exterior sides of each PCB. The exterior strips are offset by 5 mm from the interior strips to better define the field everywhere within the field cage. The cathode surface at the bottom of the field cage is constructed from graphite-coated aluminum honeycomb that is bonded to the field cage walls by epoxy. The top of the field cage is open to the wire and pad plane region, but the walls are bonded by epoxy at the top to an aluminum flange that mates via o-ring seals to a polycarbonate spacer ring and the top plate to make the field cage a gas-tight volume. This rigid and gas tight structure has a thin (4 µm) PPTA upstream beam-entrance window (6 cm wide x 7 cm high), and a larger (39cm x 81cm) and thicker (125 µm) polyamide exit window that allows passage of light charged particles and heavy ions with minimal energy loss to ancillary detectors downstream. The gas tight construction allows the possibility of using a counter gas within the detection volume inside the field cage that is different from the insulation gas in the volume between the field cage and the surrounding enclosure. Aluminum electrode surfaces were evaporated on the entrance and exit windows. These and the copper electrodes on the PCBs provide the electric field that drifts the ionized electrons to the wire planes. GARFIELD [8] simulations for the electric field in the field cage geometry confirmed its uniformity except for trajectories that pass within 1 cm of the electrodes.

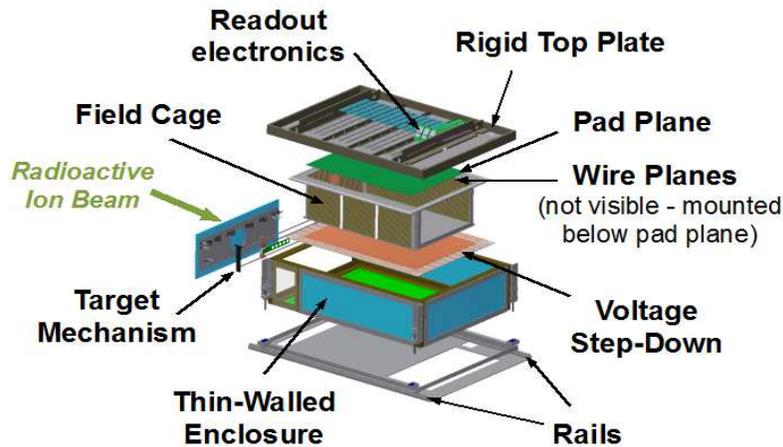

**Figure 2: Exploded view of SπRIT TPC design:** *An exploded view of the design shows the primary components of the TPC. More information is available at ref. [7].*

The bottom plate of the TPC gas enclosure lies about 2 cm below the cathode and is constructed by laminating a 6 mm polycarbonate sheet on top of an aluminum plate. Conducting epoxy was used to paint a conducting surface on the polycarbonate sheet that mirrors the cathode immediately above and was connected to it electrically. This surface was stepped down to ground by a concentric series of copper rings spaced 1.4 cm apart. High voltage was applied to the Cathode through a 10 MOhm resistor. This, combined with the 9 nF capacitance of the conductive sheet - the bottom plate, effectively filters out high-frequency noise on the cathode.

## 2.1   Wire planes and pad plane

The arrangements of the pad and wire planes for the SπRIT TPC resemble that of the EoS TPC [6]. The pad plane was tiled with 12 mm x 8 mm pads arranged in 112 rows and 108 columns; the columns and the long dimension of the pads are aligned with the beam momentum. If one defines a Cartesian coordinate system wherein the beam lies initially along the z-axis and the magnetic field lies along the y-axis, the pad plane lies parallel to the x-z plane (transverse to the magnetic field). Each pad in the pad plane resides on a six-layer PCB, with ground layers separating the charge-sensitive pads, traces, and readout connections (see Figure 3 for a cross-sectional view). The total capacitance of the pad to the environment is about 15 pF. Because of its large size, the pad plane was fabricated from four separate halogen-free G10 PCBs, which were held flat by a precision-machined vacuum table during gluing. Laser position measurements, taken after the PCBs were glued, indicate that the separation between pad plane and anode wires is constant to within 125 µm.

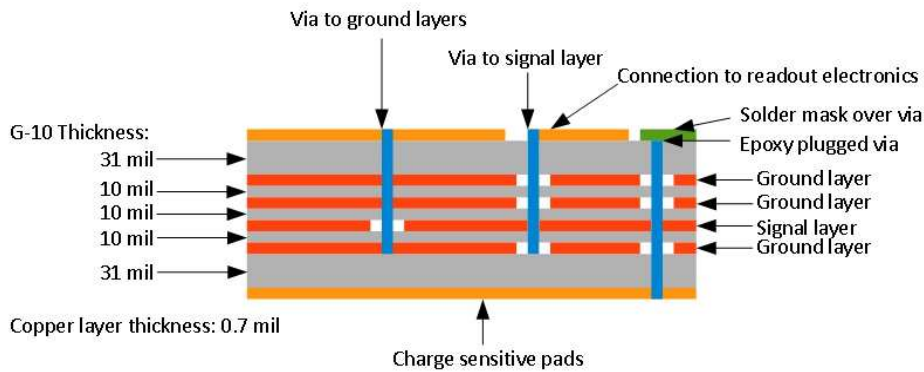

Figure 3: **Pad plane circuit board layers:** *Cross sectional view of the six-layer pad plane circuit board, illustrating the G10 thickness, the ground layers that shield the signal layer from the charge sensitive pads and a typical connection from the pad through the PCB to the readout electronics. This latter connection goes up through one ground plane, then horizontal along the signal layer, and finally up again to the readout electronics.*

Three wire planes are mounted to the bottom of the top plate just below the pad plane, with wires that run parallel to the x-axis. Table 2 lists the specifications of each wire plane. The anode plane, situated 4 mm below the pad plane, was constructed with 20 µm diameter gold plated Tungsten wires, at a spacing of 4 mm. Each pad has 3 anode wires below it, with the center wire bisecting the pad. The ground plane, situated 4 mm below the anode plane, was constructed with 76 µm diameter Beryllium Copper wires, spaced 1 mm apart with every 4$^{th}$ wire directly below an anode wire. The gating grid was situated 6 mm below the ground grid with a spacing of 1mm between 76 µm diameter Beryllium Copper wires and so that every 4$^{th}$ wire also lies directly below an anode wire. To facilitate possible replacement of broken wires, each wire plane was divided into 14 sections with separate circuit boards holding 26 anode wires or 104 ground or gating grid wires. These sections can be independently removed to allow access to the wire planes or pad planes underneath.

| Wire-Plane Specifications | | | | | | | |
|---|---|---|---|---|---|---|---|
| Plane | Material | Diameter (µm) | Pitch (mm) | Distance to pad plane (mm) | Tension (N) | Voltage (V) | No. of wires |
| Anode | Au-plated W | 20 | 4 | 4 | 0.5 | 1420 | 364 |
| Ground | BeCu | 76 | 1 | 8 | 1.2 | 0 | 1456 |
| Gating | BeCu | 76 | 1 | 14 | 1.2 | -110±70 | 1456 |

**Table 2: Wire-plane specifications**

Each anode wire was connected to the high voltage by a 10 MOhm resistor and to an external ground with a 1 nF capacitor. The high voltage and ground are connected separately in each section of anode wires, allowing the voltage of each section to be varied independently. This also allows signal readout (through the capacitors) of a single section of wires. The ground wires are all connected to a shared bus (on both ends of the wires for redundancy). Connecting a pulser to this ground wire plane, allows one to induce signals on the anode wires and the pad plane. This also allows one to cross-calibrate the gains of the electronics.

The gating grid is used to block electrons and ions from passing between the anode region and the main detector volume and to keep the detector off until a desired collision occurs. This prevents positive ions from drifting from the anode into the drift region and prevents the associated excessive space charge buildup and electric field distortions. In normal operation, alternating wires of the gating grid are biased to -180 V and -40 V when the gate is closed or -110 V when the gate is open. These bias voltages are supplied by the positive and negative conductors of two 4 Ohm stripline transmission lines. These transmission lines are electrically connected in parallel but situated at opposite ends of the gating grid wires. By connecting these low impedance lines to two high voltage switches and closing these switches, one shorts the alternating -180V and -40 V wires of the gating grid and discharges them, bringing all wires quickly to a common voltage of -110 V. The low impedance of these transmission lines, allows the gating grid to be discharged with a characteristic decay time constant of about 55 ns. In about 200 ns, the gating grid is essentially discharged, letting electrons to pass through it and multiply on the anode wires.

## 2.2 Electronics readout

Signals from the pads in the SπRIT TPC pad plane are amplified, digitized and read out by the newly developed Generic Electronics for TPCs (GET) described in Ref. [9]. Signals from the pads are sent by short cables to the inputs of the AGET chips, four of which are mounted on each AsAd motherboard. Each AGET is an application specific integrated circuit that services 64 pads, contains a Preamp (PA), a Switched Capacitor Array (SCA) (with a maximum of 512 time buckets at 1 to 100 MHz) and has multiplexer and inspection functions. Readout of these AsAd boards is performed using a Concentration Board (CoBo), which resides in a µTCA crate. Each CoBo can service up to 4 AsAds or up to 1024 pads. A total of twelve CoBo's are required to read out the SπRIT TPC.

The gain and filter of each AGET can be configured. The gain/channel settings are 0.12, 0.24, 1.0 and 10pC. The peaking times of the shaping amplifiers can be 69, 117, 232, 501, 720 or 1014 ns. Currently, the SπRIT TPC uses the 0.12pC setting, which is similar to the gain setting for the Front End Electronics (FEE) cards of the STAR TPC [10]. Each channel has a discriminator which allows for selective readout of live channels. The 12 bit ADC's on the AsAd boards deliver an effective 10.5 bit resolution.

Each CoBo contains a Xilinx Virtex 5 FPGA chip, which is coupled to a fast memory with a double buffer architecture. Currently, each CoBo card configures its four AsAd cards, collects their ADC outputs and performs data reduction, time stamping and formatting functions before transferring the data at 1Gb/s to the 10Gb/s µTCA switch. The GET system uses Gigabit Ethernet via TCP/IP and embedded LINUX and VxWorks. It is expected that the maximum event rate for this system can be 500 events/s. The anticipated data rate for the first series of SπRIT TPC experiments is less than 100 events/s.

## 3 Summary

A Time-Projection Chamber called the SAMURAI Pion-Reconstruction and Ion-Tracker (SπRIT) has recently been constructed at Michigan State University and shipped to RIKEN. The SπRIT TPC will be placed in the magnet chamber of the SAMURAI Spectrometer at the Radioactive Isotope Beam Factory (RIBF) at RIKEN to measure yield ratios for pions and other light isospin multiplets produced in central collisions of neutron-rich and neutron-deficit heavy ions. The incorporation of thin windows in the design allows passage of charged particles to ancillary detectors downstream of the TPC. The SπRIT TPC employs the newly developed GET TPC readout electronics, which allows fast readout of the TPC with minimal associated deadtime.

## 4 Acknowledgment

This material is based on work supported by the Department of Energy under Grant No. DE-SC0004835, Japanese Grant in Aid award, National Science Foundation under Grant No. PHY-1102511 and the National Research Foundation of Korea under grant No. 2012M7A1A2055596. We also acknowledge travel support provided by DOE Funding Opportunity Award DE-PS02-08ER08-10.